# Removal of Communication Gap

Z. Ahmed[ab]. S.Ganti[b]

[a] *Vienna University of Technology, Austria*
[b] *Blekinge Institute of Technology, Sweden*

**Abstract**

This research is about an online forum designed and developed to improve the communication process between alumni, new, old and upcoming students. In this research paper we present targeted problems, designed architecture, used technologies in development and final end product in detail.

**1. Introduction**

BTH – Student Forum is a web application architect and developed to improve the communication between alumni, senior, junior and upcoming students of the Blekinge Institute of Technology(BTH) Sweden [1]. By using this forum students can share current circumstances and aware new national and international students about the present weather, educational, environmental, social, economical and living conditions.

Main clients of this online web application are the students of Blekinge Tekniska Högskola. The requirements have been finalized according to the needs of the students, requirements are .i.e.,

1. A Dynamic Web Application with name BTH – Student Forum.
2. Able to browse using every recommended browser like Microsoft Internet Explorer [2], Mozilla [3] etc.
3. Able to upload information.
4. Secure and authenticated user registration and login process.
5. Open discussion section for online user communication,
6. User can make and update his personal profile.
7. User can view the already posted message's headlines, messages in detail , contacts of the posted message's user, replies to a particular message, contacts of the posted reply's user
8. User can post a message.
9. User can post the reply to message.
10. User can communicate with already login users with open discussion section.
11. User can unsubscribe his self.
12. Proper session will be maintained.
13. Complete record of users, messages3 and

replies should be stored and maintained in database.

## 2. BTH Forum – Design Implementation

### 2.1. Architecture

The architecture consists of three layers .i.e., Presentation Layer, Business Layer, DB Layer and one repository called Database as shown in Fig. 1.

#### 2.1.1. Presentation Layer
This layer is the main front end of the applications;

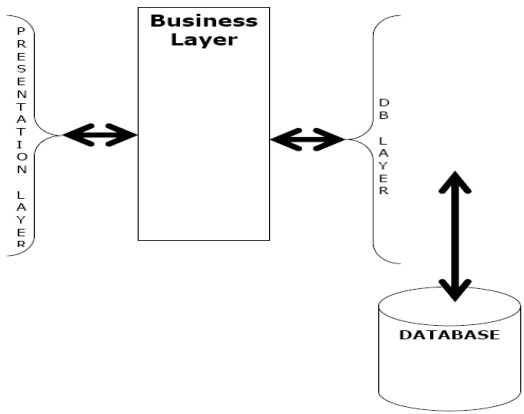

Fig. 1. Architecture

the job of this layer is to provide information to the user, and allow user to interact with software by dong possible jobs. Moreover this layer takes requests from users and forward to Business Layer for further internal processing.

#### 2.1.2. Business Layer
This layer is actually the back bone of the product; the job of this layer is to take and send the values from the presentation layer and make transactions using DB Layer.

#### 2.1.3. DB Layer
This layer is actually used to directly interact with the database; the job of this layer is to maintain connections and transactions to the database.

#### 2.1.4. Database
This is the database of the project; this consists of all the tables and their relations, more over this contains all the data belongs to the application.

### 2.2. Internal Work Flow

The proposed and designed internal work flow of the project as shown in Fig. 2 starts with the initialization of main web Server, then application is supposed to run using internet browser. If the user is a new user then application allows him to first register himself then login, and if the user is already registered then application allows user to login by entering valid user name and password.

After Logging in to the application, now user is allowed to view already posted messages, post new messages, chat with other online users using open online discussion section, change his profile settings, logout and unsubscribe.

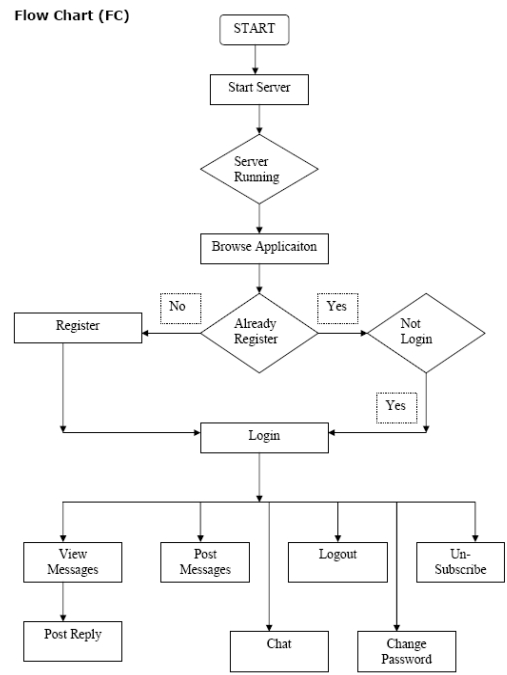

Fig. 2. Internal Work Flow

### 2.3. Class Design

The whole application implementation class designs are divided into there are three different class categories .i.e., JSP, Servlets and Beans.

#### 2.3.1. JSP Classes
JSP classes are to handle the application presentation layer. The application's user interface is designed using JSP with embedded Java Script [4], XHTML, HTML and XML [5], as show in Fig. 3.

Following nine are the JSP classes (web pages) .i.e.,
1. Home.jsp : Home page of the application.
2. register.jsp : Page to register the user in to the system.
3. signin.jsp : Page to sign in the user in to the forum
4. embers.jsp: Page to identify member.
5. essage.jsp: Page to represent the messages posted by the users.
6. reply.jsp: Page to represent the reply posted against the messages by the users
7. chat.jsp: Page to provide online chat for online users.
8. forward.jsp: Page to forward the messages
9. error.jsp: Page to perform the exception handling explicitly.

### 2.3.2. Servlets

Servlets are used to implement the business layer of the layer architecture as shown in Fig. 4. All the business logic has been implemented in the Servlets (based java classes). The jobs of business layer (Servlets) are to take user requests and process them via Beans explained latter in section 3.3.3. Following are 11 Servlets classes .i.e.,
1. RegisterServlet.java: This Servlet is used to make the registration process, takes user information from register.jsp and then register the user.
2. LoginServlet.java: This Servlet is used to implement login process, take inputs (user name and password) from signin.jsp and login the user.
3. LogoutServlet.java: This Servlet is used to log out the user.
4. PasswordServlet.java: This Servlet is to allow user to change password.
5. MessageListServlet.java: This Servlet is to maintain the message list posted by users.
6.

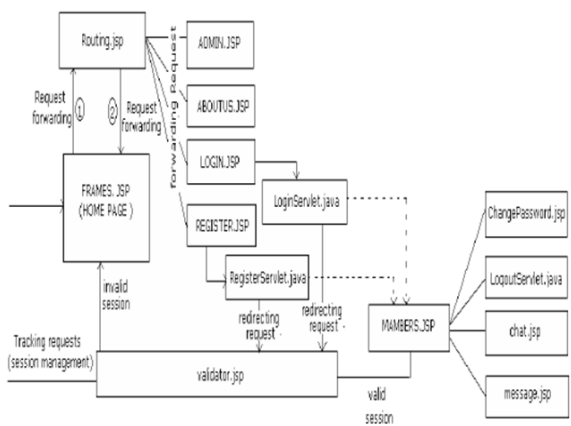

Fig. 3. JSP

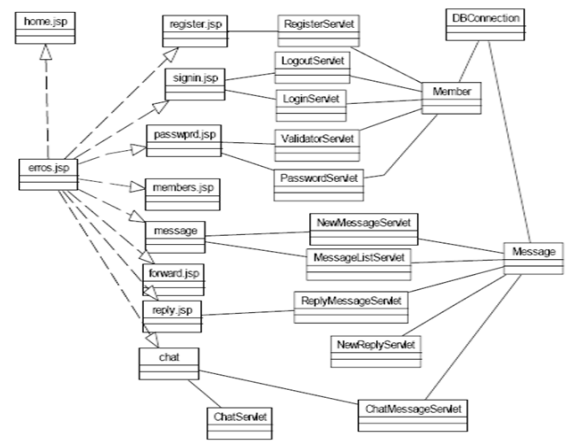

Fig. 4. Servlets

7. NewMessageServlet.java: This Servlet is to update message list by adding the newly received message.

8. NewReplyServlet.java: This Servlet is to update the reply list by adding the newly received reply from the user.

9. eplyMessageServlet.java: This Servlet is to maintain the reply message list posted by the users.

10. ChatMessageServlet.java: This Servlet is to maintain the communication between the users.

11. ChatServlet.java: This Servlet is to update user chat with user sent messages.

12. ValidatorServlet.java: This Servlet is to perform the validation process.

### 2.3.3. Beans

Beans are the wrapper classes. The task of these classes is to form the Database layer. Beans are used to communicate with the database by connecting database and manage data transactions and manipulation. Beans are also called the instances of the relations of the data base, for each relation of data base there is a wrapper class and each relation is represented by java object.

Following are three Bean classes .i.e.,

1. DBConnection.java: This bean class is to maintain the connection to the data base.
2. Member.java: This Bean class is to maintain data manipulation and transactions like registration and sing in etc.
3. Message.java: This Bean class is to maintain the data manipulation and transactions consisting of messages and replies.

### 2.4. Database

The designed entity relationship diagram is consists of three main entities .i.e., forumMember, forumMessages, forumMemberCategory and two relationships .i.e., Has and Have as shown in Fig. 5.

#### 2.4.1. forumMember

forumMember relation is consists of five attributes .i.e., memberName, Password, handle, memberDOB and memberDOJ. MemberName is to store user name, password is to store user password, memberDOB is to store user's date of birth, memberDOJ is to store the member's date of joining and handle is store the current status.

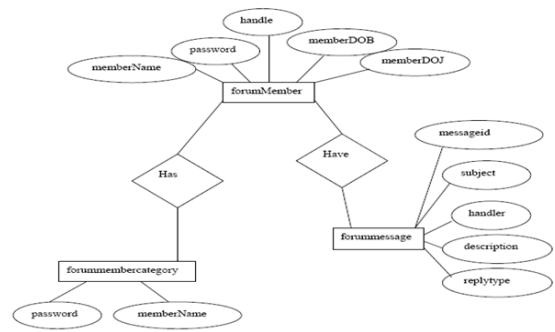

Fig. 5. ERD

#### 2.4.2. forumMember

forumMessages relation is consists of five attributes .i.e., messageid, subject, handler, description and replytype. Messageid is to store the each message's unique id, subject is to store the running subject of message, hanlder is the status of message, description is to store the message data and replytype is to store the type reply.

#### 2.4.3. forumMemberCategotry

forumMemberCategory relation is consists of two attributes .i.e., password and memberName. Password is to store the admin member password and memberName is the admin password name.

#### 3.4.4. Has

Has is one to one relationship between forumMember and forumMemberCategory.

*2.4.4. Have*

Have is many to many relationship between forumMember and forumMessages.

*2.4. Web Page Map*

This Web page Map is seven different web pages .i.e., Main Page, Introduction Page, Registration Page, Sign In Page, Contact Us Page, About US Page, Chat and Message as shown in Fig. 6.

User can simply browse Introduction, Registration, Sign In, Contact Us and About Us

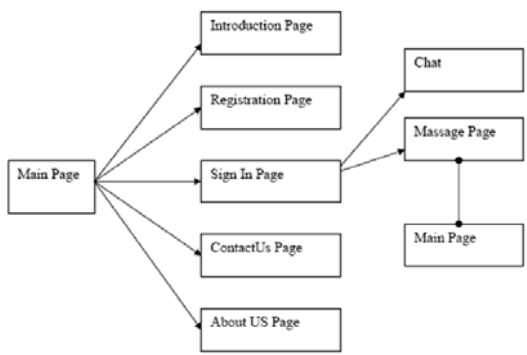

Fig. 6. Web Page map

web pages using main page but can only access chat and message page after login in to the application.

## 3. Technologies Involved

Programming languages, tools and technologies which are used during the process of development are .i.e., JAVA (Servlets, JAVA Script, JSP) [4], HTML [5], XHTML [5], XML [5], Tomcat Server [6], MySql [7], UML [8] and Microsoft office [9].

## 4. BTH – Student Forum Version 1

Implementing designed designs using involved tool and technologies BTH-Studetn Forum's first version has been implemented as shown in Fig 7, 8 and 9.

## 5. Conclusion

This research paper is about a research project to improve the process of communication by reducing the gaps between alumni, old, junior and upcoming student called BTH-Student Forum.
In this research paper we provided all project based information containing all the information about requirements specifications, designs of application inclosing architecture, class, internal work flow, database and web page map designs. Moreover, in the end, we have also provided the implemented version of BTH-Student Forum.

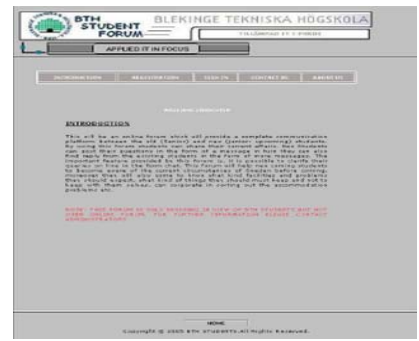

Fig. 7. Web Page map

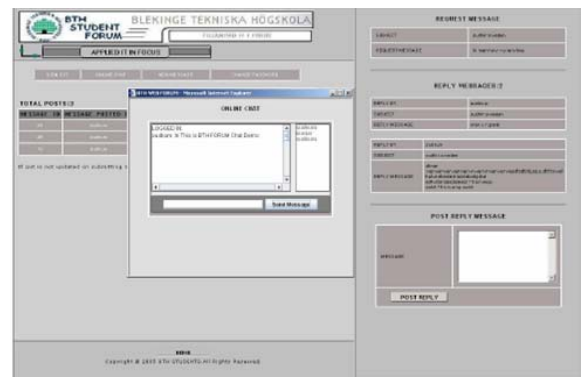

Fig. 8. Message and Chat

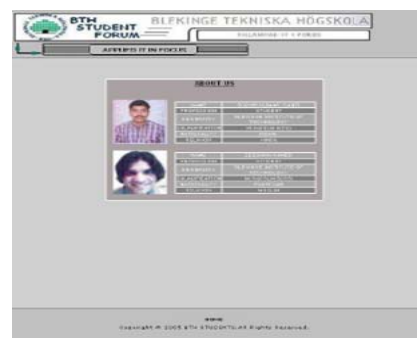

Fig. 9. About Us